\documentclass[twocolumn,amsmath,showpacs,amsfonts,aps,prc,floatfix]{revtex4}
\usepackage{graphicx}
\usepackage{bm}
\newcommand{\dn}{\frac{dN^\phi}{dy}} 
\newcommand{\pt}{<p_T^\phi>} 
\begin{document}

\title{Centrality dependence of elliptic flow and QGP viscosity} 
 
\author{A. K. Chaudhuri}
\email[E-mail:]{akc@veccal.ernet.in}
\affiliation{Variable Energy Cyclotron Centre, 1/AF, Bidhan Nagar, 
Kolkata 700~064, India}

\begin{abstract}

In the Israel-Stewart's theory of second order hydrodynamics, we have analysed the recent PHENIX data on charged particles elliptic flow in Au+Au collisions.
  PHENIX data demand
more viscous fluid in peripheral collisions than in central collisions.  
Over a broad range of collision centrality (0-10\%- 50-60\%), viscosity to entropy ratio ($\eta/s$) varies between 0-0.17.

  \end{abstract}


\date{\today}  

\maketitle


\section{Introduction} \label{intro}
 
 One of the important results in Au+Au collisions at RHIC is the 
large elliptic flow in non-central collisions \cite{BRAHMSwhitepaper,PHOBOSwhitepaper,PHENIXwhitepaper,STARwhitepaper}. Large elliptic flows establish that in non-central Au+Au collisions, a collective QCD matter is created. Whether the matter can be characterized as the lattice QCD \cite{lattice,Cheng:2007jq} predicted Quark-Gluon-Plasma (QGP) or not,   is still a question of debate. 
Qualitatively, elliptic flow is naturally explained in a hydrodynamical model,
rescattering of secondaries generates pressure and drives the subsequent collective motion. In non-central collisions, the reaction zone is asymmetric (almond shaped), pressure gradient
is large in one direction and small in the other. The asymmetric pressure gradient generates the elliptic flow. As the fluid evolve and
expands, asymmetry in the reaction zone decreases and  a stage arise when reaction zone become symmetric and system no longer generate elliptic flow.  
Elliptic flow is early time phenomena. It is  sensitive probe to, (i) degree of
thermalisation, (ii) transport coefficient and (iii) equation of state of the early stage of the fluid \cite{Ollitrault:1992bk,Kolb:2001qz,Hirano:2004en}.

Dissipative effects like viscosity reduce elliptic flow. The 
conversion of initial spatial anisotropy to momentum anisotropy is hindered in presence of viscosity and elliptic flow is  reduced. 
QGP viscosity is an important parameter, 
theoretical estimates of the ratio, (shear) viscosity over the entropy density, $\eta/s$ cover a wide range,   0.0-1.0. 
String theory based models (ADS/CFT) give a lower bound on viscosity of any matter $\eta/s \geq 1/4\pi$ \cite{Policastro:2001yc}. In a perturbative QCD, Arnold et al  \cite{Arnold:2000dr} estimated $\eta/s\sim$ 1.  In a SU(3) gauge theory, Meyer \cite{Meyer:2007ic} gave the upper bound $\eta/s <$1.0, and his best estimate is $\eta/s$=0.134(33) at $T=1.165T_c$.  At RHIC region, Nakamura and Sakai \cite{Nakamura:2005yf} estimated the viscosity of a hot gluon gas  as $\eta/s$=0.1-0.4. Attempts have been made to estimate QGP viscosity directly from experimental data.  Gavin and Abdel-Aziz \cite{Gavin:2006xd} proposed to measure viscosity from transverse momentum fluctuations. From the existing data on Au+Au collisions, they estimated  QGP viscosity as $\eta/s$=0.08-0.30. Experimental data on elliptic flow has also been used to estimate QGP viscosity. Elliptic flow scales with eccentricity. Departure from the scaling can be understood as due to off-equilibrium effect and utilised to estimate viscosity \cite{Drescher:2007cd} as, $\eta/s$=0.11-0.19. Experimental observation that elliptic flow scales with transverse kinetic energy is also used to estimate QGP viscosity, $\eta/s \sim$ 0.09 $\pm$ 0.015 \cite{Lacey:2006bc}, a value close to the ADS/CFT bound. From heavy quark energy loss, PHENIX collaboration \cite{Adare:2006nq} estimated QGP viscosity $\eta/s\approx$ 0.1-0.16.  Recently, from analysis of RHIC data, in a viscous hydrodynamics, upper bound to viscosity is given $\eta/s <$ 0.5 \cite{Luzum:2008cw,Song:2008hj}. Using a fully lattice based equation of state (lattice simulations for both the confined and the deconfined phase),  we have estimated QGP viscosity as, $\eta/s=0.15 \pm 0.06$ \cite{Chaudhuri:2009vx}. In a more recent analysis \cite{Chaudhuri:2009uk}, with a lattice based equation of state ( lattice simulations for the confined phase and non-interacting hadronic resonance gas for the hadron phase), viscosity was estimated as, $\eta/s=0.07\pm 0.03 \pm 0.14$. In \cite{Chaudhuri:2009uk}, estimate of viscosity was obtained from analysing STAR data on $\phi$ mesons multiplicity, mean $p_T$ and integrated $v_2$. In the same model from a study of 'scaling departure' of elliptic flow, viscosity over the entropy ratio was estimated as, $\eta/s=0.12 \pm 0.03$ \cite{Chaudhuri:2009ud}. In the present paper, we have studied the centrality dependence of elliptic flow in ideal and viscous hydrodynamics. 
From a direct fit to the PHENIX data \cite{Afanasiev:2009wq} on the charged particles elliptic flow in 0-60\% centrality collisions, we have also obtained an estimate of QGP viscosity, $\eta/s=0.115 \pm 0.005$, very close to the value obtained from the study of scaling departure of elliptic flow \cite{Chaudhuri:2009ud}. Our analysis also indicate that the PHENIX data on the centrality dependence of   differential elliptic flow require stronger viscosity in   peripheral collisions than in central collisions. Over the broad range of collision centrality (0-10\% - 50-60\%),
viscosity over the entropy ratio varies between 0 -0.17.
 

The paper is organised as follows: in section II, we briefly explain the 2nd order Israel-Stewarts theory of dissipative hydrodynamics. Equation of state,
initialisation of the fluid is described in section III. Results are discussed in section VI. The summary and conclusions are given in section V.

\section{Hydrodynamic model}

In the following, we consider a baryon free fluid with only shear viscosity.
Bulk viscosity is neglected. In the Israel-Stewart's theory of 2nd order dissipative hydrodynamics, equation of motion of the fluid is obtained by solving,
  
\begin{eqnarray}  
\partial_\mu T^{\mu\nu} & = & 0,  \label{eq3} \\
D\pi^{\mu\nu} & = & -\frac{1}{\tau_\pi} (\pi^{\mu\nu}-2\eta \nabla^{<\mu} u^{\nu>}) \nonumber \\
&-&[u^\mu\pi^{\nu\lambda}+u^\nu\pi^{\mu\lambda}]Du_\lambda. \label{eq4}
\end{eqnarray}

Eq.\ref{eq3} is the conservation equation for the energy-momentum tensor, $T^{\mu\nu}=(\varepsilon+p)u^\mu u^\nu - pg^{\mu\nu}+\pi^{\mu\nu}$, 
$\varepsilon$, $p$ and $u$ being the energy density, pressure and fluid velocity respectively. $\pi^{\mu\nu}$ is the shear stress tensor (we have neglected bulk viscosity and heat conduction). Eq.\ref{eq4} is the relaxation equation for the shear stress tensor $\pi^{\mu\nu}$.   
In Eq.\ref{eq4}, $D=u^\mu \partial_\mu$ is the convective time derivative, $\nabla^{<\mu} u^{\nu>}= \frac{1}{2}(\nabla^\mu u^\nu + \nabla^\nu u^\mu)-\frac{1}{3}  
(\partial . u) (g^{\mu\nu}-u^\mu u^\nu)$ is a symmetric traceless tensor. $\eta$ is the shear viscosity and $\tau_\pi$ is the relaxation time.  It may be mentioned that in a conformally symmetric fluid relaxation equation can contain additional terms  \cite{Song:2008si}.

Assuming boost-invariance, Eqs.\ref{eq3} and \ref{eq4}  are solved in $(\tau=\sqrt{t^2-z^2},x,y,\eta_s=\frac{1}{2}\ln\frac{t+z}{t-z})$ coordinates, with the code 
  "`AZHYDRO-KOLKATA"', developed at the Cyclotron Centre, Kolkata.
 Details of the code can be found in \cite{Chaudhuri:2008sj}. 
Within 10\% or less, AZHYDRO-KOLKATA simulation  reproduces  Song and Heinz's  \cite{Song:2008si} result for temporal evolution of momentum anisotropy $\varepsilon_p$. 

\section{Equation of state}

\begin{figure}[t]
\vspace{0.3cm} 
\center
 \resizebox{0.35\textwidth}{!}{%
  \includegraphics{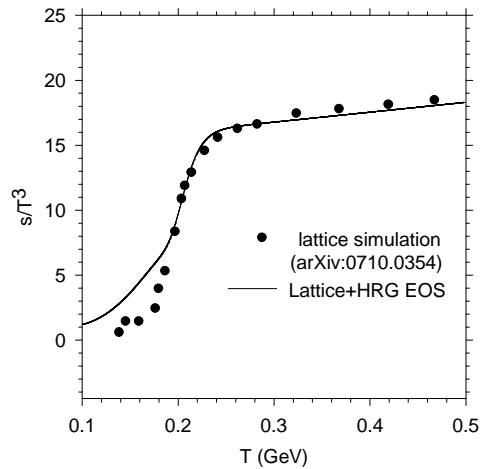}
}
\caption{Lattice simulation for entropy density is compared with the same in model EOS, lattice+HRG.
The filled circles are lattice simulation \cite{Cheng:2007jq} for $s/T^3$. The solid line is the same in model EOS (see text).}
  \label{F1}
\end{figure}

Eqs.\ref{eq3},\ref{eq4} are closed with an equation of state (EOS) $p=p(\varepsilon)$.
Lattice simulations \cite{lattice,Cheng:2007jq} indicate that the confinement-deconfinement transition is a cross over, rather than a 1st or 2nd order phase transition.   In Fig.\ref{F1}, a recent lattice simulation  \cite{Cheng:2007jq} for the  entropy density is
  shown. 
  We complement the lattice simulated EOS \cite{Cheng:2007jq} by a
  hadronic resonance gas (HRG) EOS comprising all the resonances below mass 2.5 GeV. In Fig.\ref{F1}, the solid line is the
   entropy density of the "`lattice +HRG"' EOS. The entropy density is obtained as,
     
   \begin{equation}
   s=0.5[1-tanh(x)]s_{HRG} + 0.5 [1+tanh(x)]s_{lattice}
   \end{equation}
   
\noindent where $s_{lattice}$ and $s_{HRG}$ are entropy density from lattice simulations and HRG model, $x=\frac{T-T_c}{\Delta T}$. In the present simulation, we have used cross over temperature, $T_c$=196 MeV and $\Delta T=0.1T_c$.  
 Compared to lattice simulation, entropy density in HRG drops slowly at low temperature.
It is consistent with observation in \cite{Cheng:2007jq}, that at low temperature, trace anamoly,
$\frac{\varepsilon-3p}{T^4}$ drops faster in lattice simulation than in a HRG model. It is difficult to resolve whether the discrepancy is due to failure of HRG model at lower temperature
or due to the difficulty in resolving low energy hadron spectrum on   rather coarse lattice \cite{Cheng:2007jq}.

From the entropy density,   using  the thermodynamic relations,

\begin{eqnarray}  
  p(T)&=&\int_0^T s(T^\prime) dT^\prime \label{eq2a} \\
  \varepsilon(T)&=&Ts -p \label{eq2b},
  \end{eqnarray}
  pressure and energy density can be obtained.

\section{Initialisation of the fluid}

Solution of partial differential equations (Eqs.\ref{eq3},\ref{eq4}) requires initial conditions, e.g.  transverse profile of the energy density ($\varepsilon(x,y)$), fluid velocity ($v_x(x,y),v_y(x,y)$) and shear stress tensor ($\pi^{\mu\nu}(x,y)$) at the initial time $\tau_i$. One also need to specify the viscosity ($\eta$) and the relaxation time ($\tau_\pi$). A freeze-out prescription is also needed to convert the information about fluid energy density and velocity to particle spectra and compare with experiment.

We assumed that the fluid is thermalised at $\tau_i$=0.6 fm \cite{QGP3} and the initial fluid velocity is zero, $v_x(x,y)=v_y(x,y)=0$. Initial energy density is assumed to be distributed as \cite{QGP3}

\begin{equation} \label{eq6}
\varepsilon({\bf b},x,y)=\varepsilon_i[0.75 N_{part}({\bf b},x,y) +0.25 N_{coll}({\bf b},x,y)],
\end{equation}

\noindent
where b is the impact parameter of the collision. $N_{part}$ and $N_{coll}$ are the transverse profile of the average number of  participants and average number collisions respectively, calculated in a Glauber model. Central energy density, 
$\varepsilon_i$ is a parameter and does not depend on the impact parameter of the collision.    The shear stress tensor was initialised with boost-invariant value, $\pi^{xx}=\pi^{yy}=2\eta/3\tau_i$, $\pi^{xy}$=0. For the relaxation time, we used the   Boltzmann estimate $\tau_\pi=3\eta/2p$. Finally, the freeze-out was fixed at $T_F$=150 MeV \cite{note1}.

\begin{table}[h]
\caption{\label{table1} Initial central energy density ($\varepsilon_i$) and temperature ($T_i$) of the fluid in b=0 Au+Au collisions, for different values of viscosity to entropy ratio ($\eta/s$). The predicted $\phi$ meson multiplicity
and mean $p_T$ are also noted. 
} 
  \begin{tabular}{|c|c|c|c|c|}\hline
$\eta/s$  & $\varepsilon_i$    & $T_i$  & $\frac{dN^\phi}{dy}$ & $<p_T>$\\   
& $(GeV/fm^3)$ & (MeV)  &   &(GeV) \\  \hline
0 &    $35.5\pm 5.0$ & $377.0\pm 13.7$ &7.96 &1.02\\
0.08 &  $29.1\pm 3.6$ & $359.1\pm 11.5$ &8.01&1.06 \\
0.12 & $25.6\pm 4.0$ & $348.0\pm 14.3$ &8.22 &1.11\\
0.16 & $20.8\pm 2.7$ & $330.5\pm 11.3$ &8.13 &1.17\\ \hline
\end{tabular}
\end{table}    

Assuming that throughout the evolution, viscosity to entropy ratio ($\eta/s$) remains a constant, we have simulated Au+Au collisions for four values of $\eta/s$, (i) $\eta/s$=0 (ideal fluid), (ii) $\eta/s=1/4\pi\approx$ 0.08 (ADS/CFT lower limit of viscosity), (iii)   $\eta/s$=0.12 and (iv) $\eta/s$=0.16.
Note that during the evolution, the fluid cross over from QGP phase to hadronic phase.  Constant $\eta/s$ can be thought over as a space-time averaged $\eta/s$.
As the entropy density of the QGP phase is more than that of a HRG (see Fig.\ref{F1}),  constancy of $\eta/s$ during the evolution implicitly assume $\eta_{QGP} > \eta_{HRG}$. It also fixes the temperature dependence of $\eta$,
Variation of $\eta$ with temperature is the same as that of the entropy density. 

 \begin{figure}[t]
 \center
 \resizebox{0.4\textwidth}{!}{%
  \includegraphics{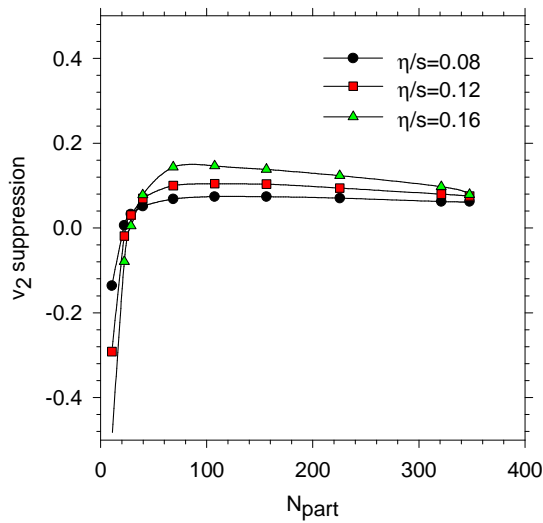} 
}
\caption{(color online) Viscous suppression of elliptic flow in Au+Au collisions. The solid, dashed and mid-dashed lines are fractional suppression of elliptic flow in viscous fluid evolution with $\eta/s$=0.08, 0.12 and 0.16 respectively.}
 \label{F2}
\end{figure}

Only parameter left to be initialised is the central energy density $\varepsilon_i$.
We fix   $\varepsilon_i$ such that entropy at the freeze-out is the same either in ideal or in viscous evolution.
This is done by reproducing the STAR data \cite{Abelev:2007rw} on $\phi$ meson multiplicity in 0-5\% Au+Au collisions. In the present work, we have neglected resonance contribution. We choose $\phi$ meson multiplicity as they are not affected by resonance decays.
In table.\ref{table1},   initial central energy density ($\varepsilon_i$) and temperature ($T_i$) required to fit STAR data on $\phi$ meson multiplicity in 0-5\% Au+Au collisions are noted. The error in $\varepsilon_i$ or in $T_i$ corresponds to statistical and systematic uncertainty in STAR measurements \cite{Abelev:2007rw}. In viscous fluid evolution, entropy is generated. More viscous is the fluid, more is the entropy generation. As a consequence, viscous fluid requires less initial energy density (or temperature) than an ideal fluid.
For example, compared to ideal fluid, in minimally viscous ($\eta/s$=0.08) fluid, initial energy density is reduced by $\sim$ 18\%. In fluid with viscosity $\eta/s$=0.16, reduction is even more, $\sim$40\%. The predicted central values of $\phi$ meson multiplicities and mean $p_T$  are also shown in table.\ref{table1}. They should be compared with STAR measurements \cite{Abelev:2007rw},   ${\dn}^{ex}=7.95 \pm 0.74$, and $\pt=0.977\pm 0.064$ (statistical and systematic error included).   Experimental data on $\phi$ meson multiplicity and mean $p_T$ in 0-5\% centrality Au+Au collisions are simultaneously explained only for $\eta/s \leq$0.12. 
We have not shown it here, but hydrodynamic evolution of the fluid initialised as stated here, reproduce experimental $p_T$ spectra of pions and kaons at $p_T >$1 GeV ($p_T < 1$ GeV part of the spectra is under predicted due to neglect of resonance contribution).  It also reproduces the  $p_T$ spectra of $\phi$ mesons. The proton spectra however is underestimated by a factor of $\sim$ 2.

 \begin{figure}[t]
 \center
 \resizebox{0.4\textwidth}{!}{%
  \includegraphics{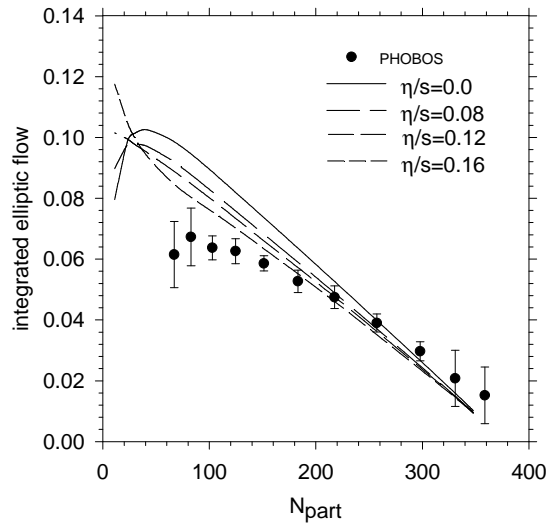} 
}
\caption{(color online) The symbols are PHOBOS data on the centrality dependence of charged particles integrated $v_2$. The solid, dashed, medium-dashed and short-dashed lines are simulated flow with $\eta/s$=0, 0.08, 0.12 and 0.16 respectively.}
 \label{F3}
\end{figure}

\begin{figure}[t]
 \center
 \resizebox{0.4\textwidth}{!}{%
  \includegraphics{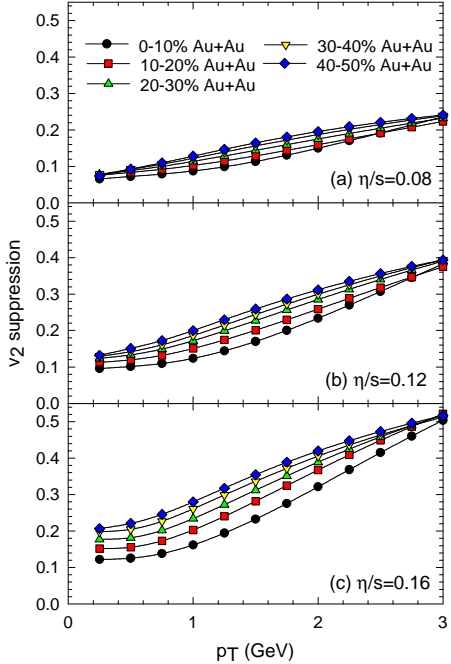} 
}
\caption{(color online) Fractional suppression of differential $v_2$ in Au+Au collisions in the 0-60\% centrality ranges of collisions.}
\label{F4}
\end{figure}

\begin{figure}[t]
 \center
 \resizebox{0.4\textwidth}{!}{%
  \includegraphics{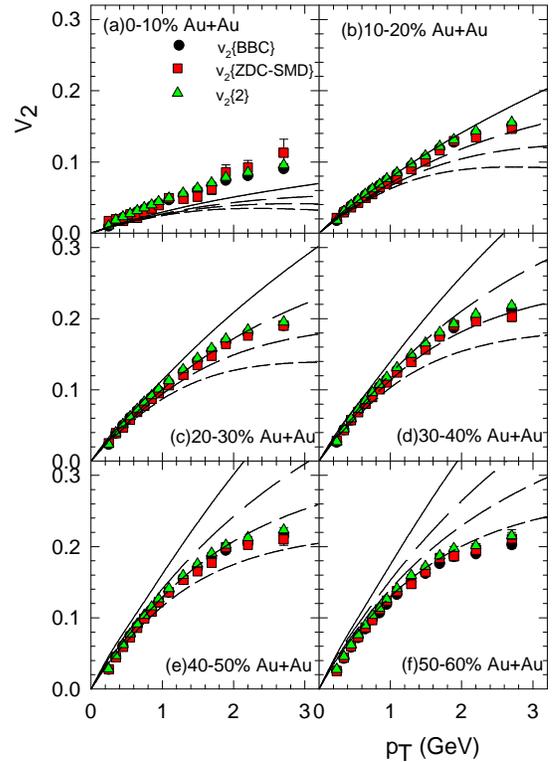} 
}
\caption{(color online) 
 In six panels, PHENIX measurements for elliptic flow in 0-10\%, 10-20\%, 20-30\%, 30-40\%, 40-50\% and 50-60\% Au+Au collisions are shown. The solid, dashed, medium dashed and short dashed lines are elliptic flow in hydrodynamic simulations with $\eta/s$=0, 0.08, 0.12 and 0.16 respectively.}
\label{F5}
\end{figure}

\begin{figure}[t]
 \center
 \resizebox{0.4\textwidth}{!}{%
  \includegraphics{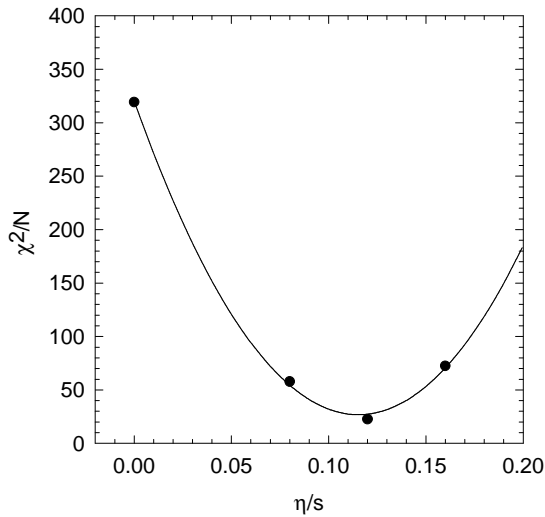} 
}
\caption{ 
  $\chi^2/N$ for the PHENIX measurements of $v_2\{2\}$ in 0-60\%  Au+Au collisions, as a function of $\eta/s$ are shown. The solid line is a  fit to the $\chi^2/N$ values by a parabola.}  
\label{F6}
\end{figure}

\section{Elliptic flow in ideal and viscous hydrodynamics}

\subsection{Centrality dependence of integrated $v_2$}

Let us first study the effect of viscosity on $p_T$ integrated elliptic  flow. 
In Fig.\ref{F2}, fractional decrease in charged particles (pions, kaons and protons)   elliptic flow, $(1-v_2^{vis}/v_2^{id})$, as a function of collision centrality is studied.
As mentioned earlier, we have neglected resonance decays. Resonance contribution reduces elliptic flow, mostly for low mass particles at low $p_T$ \cite{Hirano:2002ds}, e.g. in the $p_T$ range, $0 \leq p_T \leq 1$ GeV , $v_2$ for pions is reduced by $\sim$ 0-30\%. At $p_T >$ 1 GeV, resonance contribution to $v_2$ is negligible \cite{Hirano:2002ds}. Neglect of resonance contribution will increase integrated $v_2$, however, the effect will be    minimised in the ratio $v_2^{vis}/v_2^{id}$. 
   In collisions with $N_{part} \geq 100$, integrated $v_2$ is suppressed by $\sim$5\%, 10\% and 15\% for viscosity $\eta/s$=0.08, 0.12 and 0.16 respectively. There is also an indication that in peripheral collisions, $v_2$ is more suppressed than in central collisions. In very peripheral collisions, $N_{part} <$100, $v_2$ in viscous fluid is more than that in ideal fluid. However, in very peripheral collisions, applicability of hydrodynamics is questionable  \cite{QGP3}. We donot think that present simulations are reliable for peripheral ($N_{part} <$100) collisions. 
 
In Fig.\ref{F3}, we have compared simulated (integrated) elliptic flow with
PHOBOS measurements  \cite{Back:2004mh} for charged particles elliptic flow. In Fig.\ref{F3}, the solid, dashed, medium-dashed and short-dashed lines are the  charged particles elliptic flow in  the present simulation, with
$\eta/s$=0, 0.08, 0.12 and 0.16 respectively. Evidently, PHOBOS data on the centrality dependence of $v_2$ prefer viscous fluid rather than ideal fluid. However, one also note that   in   central collisions $N_{part} >250$, simulated $v_2$ in ideal fluid evolution  are more close to the experimental data than in viscous evolution. For $N_{part} >$ 250, data are  better explained in ideal dynamics than in viscous dynamics. Data at $N_{part} < 250$ on the other hand are better explained in viscous evolution than in ideal evolution.

\subsection{Centrality dependence of differential $v_2$}
 
Viscous suppression of differential $v_2$ is studied in Fig.\ref{F4}. 
In Fig.\ref{F4}, in three panels, for $\eta/s$=0.08, 0.12 and 0.16, fractional decrease in  $v_2$  in  0-10\%, 10-20\%, 20-30\%, 30-40\% and 40-50\%  Au+Au collisions are shown. 
In all the centrality ranges of collisions, viscous suppression is more at high $p_T$ than at low $p_T$. Suppression is also centrality dependent, more in peripheral collisions than in central collisions.  
For example, for $\eta/s$=0.12, at a fixed $p_T\approx$1 GeV, compared to a central collisions, in peripheral collisions, $v_2$ suppression is increased by a factor of $\sim$ 2.  
At large $p_T$ however, viscous suppression tends to saturate for all collision centrality.

In Fig.\ref{F5}, we have compared the presently simulated $v_2$ with PHENIX measurements \cite{Afanasiev:2009wq}. In Fig.\ref{F5}  colored symbols are PHENIX measurements \cite{Afanasiev:2009wq} for charged particles elliptic flow in 0-10\%, 10-20\%, 20-30\%, 30-40\%, 40-50\% and 50-60\%    Au+Au collisions. PHENIX collaboration measured charged particles $v_2$ upto $p_T\approx$ 8 GeV. In Fig.\ref{F5},   measurements upto $p_T$=3 GeV are shown only. Hydrodynamic models are not well suited for high $p_T$ particles.
 In order to study non-flow
effects that are not correlated with the reaction plane, as well as fluctuations of $v_2$, PHENIX collaboration obtained $v_2$ from two independent analysis,
(i) event plane method from two independent subdetectors, $v_2\{BBC\}$ and 
$v_2 \{ZDC-SMD\}$ and (ii) two particle cummulant $v_2\{2\}$.  $v_2\{2\}$ from two particle cummulant and $v_2\{BBC\}$ or $v_2\{ZDC-BBC\}$ from event plane methods agree within the systematic error.   It may also be mentioned here that   $v_2\{2\}$ in PHENIX is lower than that $v_2\{2\}$in STAR measurements,  but they agree within the systematic error. All the three measurements of $v_2$ are shown in Fig.\ref{F5}.
 
In Fig.\ref{F5},   the solid, dashed, medium dashed and short dashed lines are simulated elliptic flow in fluid evolution with (i) $\eta/s$=0 (ideal fluid), (ii) $\eta/s$=0.08, (iii) $\eta/s$=0.12, and (iv) $\eta/s$=0.16, respectively.  
Comparison of simulated elliptic flow with PHENIX measurements indicate that  in central (0-10\%) collisions, hydrodynamic evolution produces less $v_2$ than in experiment. For example, at $p_T\approx$1 GeV,
ideal fluid evolution underestimate of PHENIX measurement of $v_2 \{ZDC-SMD\}$   by $\sim$40\%.  Viscosity reduces $v_2$ and in viscous evolution $v_2$ is even more under predicted. That in central collisions, ideal hydrodynamic model,
with Glauber model initial condition produces less $v_2$ than in experiment is well known (e.g. see Fig.9 of Ref.\cite{Kolb:2001qz}). 
In a hydrodynamic model, elliptic flow depends on the initial spatial eccentricity.
In a Glauber model, initial eccentricity in  central collisions is small and elliptic flow is not developed. Initial spatial eccentricity is comparatively large in Color Glass Condensate (CGC) model initial conditions \cite{Hirano:2009ah}.
  Hydrodynamic with  CGC model initial condition  correctly predict elliptic flow in central collisions \cite{Hirano:2009ah}. Apparently CGC initial condition is better suited for central Au+Au collisions. However, since a hydrodynamic model have quite a large number of parameters, e.g. initial time, initial energy density, equation of state, freeze-out temperature etc, unless a detailed study is made, it is difficult to conclude that hydrodynamic models with Glauber model initial condition do not correctly predict $v_2$ in central collisions.  

In mid-central or in peripheral collisions, elliptic flow is over predicted in ideal fluid evolution. Data are better explained in viscous fluid evolution.   To obtain a quantitative idea about the fit obtained to the PHENIX data
by the    hydrodynamic simulations,  we have computed
 $\chi^2/N$,  

\begin{equation} \label{eq5}
\chi^2/N = \frac{1}{N} \sum_{i=1}^{i=N} \frac{(EX(i)-TH(i))^2}{ERR(i)^2}.
\end{equation}

\noindent where $EX(i)$ and $ERR(i)$ are  the PHENIX measurements for $v_2\{2\}$ and its error, $TH(i)$ is the hydrodynamic simulations for $v_2$. 
As mentioned earlier, we have neglected resonance production. Resonances contribute mainly at low $p_T$, $p_T <$ 1 GeV  \cite{Hirano:2002ds}. To remove the uncertainty due to neglect of resonance contribution, in the $\chi^2/N$  computation, we have included data only the in the $p_T$ range 1 GeV $\leq p_T \leq 3$ GeV. In table.\ref{table2}, the $\chi^2/N$ values for different centrality ranges of collisions are noted. One note that 
in most central (0-10\%) collisions, the minimum value is obtained in ideal fluid evolution. More peripheral collisions prefer viscous fluid. 
It is also be noted that the minimum $\chi^2/N$ in 0-10\% centrality collision is   factor of 2 or more larger than minimum $\chi^2/N$ obtained in other collision centrality. Compared to peripheral collisions, elliptic flow in very central collisions are not well reproduced in a hydrodynamic model.

$\chi^2/N$ for the combined data sets as a function of $\eta/s$ is shown in Fig.\ref{F6}.    The solid line in Fig.\ref{F6} is a parabolic fit to $\chi^2/N$. Best fit to the PHENIX combined data sets is obtained for $\eta/s=0.115 \pm 0.005$. The   estimate is well within the upper bound of $\eta/s$ obtained in \cite{Luzum:2008cw,Song:2008hj}. The result is very close to the estimate   $\eta/s=0.12 \pm 0.03$, also obtained in the same model,      from the study of scaling violation of elliptic flow \cite{Chaudhuri:2009ud}. It also agree with the estimate $\eta/s$=0.11-0.19   \cite{Drescher:2007cd},  $\eta/s=0.09\pm 0.015$  \cite{Lacey:2006bc} also obtained from the analysis of elliptic flow data.

PHENIX measurements of charged particle elliptic flow in  0-60\% Au+Au collisions,
are best explained with viscosity to entropy ratio $\eta/s\approx$0.115. However, as shown in Fig.\ref{F6}, the minimum $(\chi^2/N)_{min}\approx$18, is comparatively large (a good fit to data demand $(\chi^2/N)_{min}\sim$1 ). Comparatively large value $(\chi^2/N)_{min}\approx$18, indicate that hydrodynamic model with a fixed viscosity to entropy ratio do not predict 'accurately' centrality dependence of elliptic flow. Indeed, one observe from  table.\ref{table2} that in more peripheral collisions, PHENIX   data demand more viscous    fluid. For example, in 0-10\% Au+Au collisions, minimum $\chi^2/N$ is obtain in ideal fluid ($\eta/s$=0) evolution. In 10-20\% and 20-30\% collisions,  $\chi^2/N$ is minimum for $\eta/s$=0.08. In  30-40\%, 40-50\% centrality collisions, minimum $\chi^2/N$ is obtained for $\eta/s$=0.12. In 50-60\% centrality collisions, minimum $\chi^2/N$ is obtained at still higher value, $\eta/s$=0.16. Apparently, present analysis suggests that more viscous fluid is produced in peripheral   Au+Au collisions  than  in central collisions.  To obtain the centrality dependence of $\eta/s$, in each collision centrality, $\chi^2/N$ as a function of $\eta/s$ are fitted by a parabola. From the minimum of the parabola,
best fitted $\eta/s$ is obtained. In table.\ref{table3}, best fitted $\eta/s$ as a function of collision centrality are noted.   $\eta/s$ smoothly increases as the collision become more and more peripheral. While in central collisions, a nearly perfect fluid is produced, more viscous fluid is produced in peripheral collisions. 

\begin{table}[t] 
\caption{\label{table2} $\chi^2/N$ for PHENIX $v_2\{2\}$ in 0-60\% Au+Au collisions,   in fluid evolution with viscosity to entropy ratio, $\eta/s$=0, 0.08,0.12 and 0.16. } 
  \begin{tabular}{|c|c|c|c|c|}\hline
  & \multicolumn{4}{|c|} {$\chi^2/N$}  \\ \hline \
coll. centrality  & $\eta/s=0.0$ & $\eta/s$=0.08 & $\eta/s$=0.12 &$\eta/s$=0.16   \\ \hline 
\hline
 0-10\% & 11.79 &17.87 &21.51 &25.27 \\  \hline
10-20\% & 19.48 &5.78  &33.61 &93.77 \\ \hline
20-30\% & 215.21 &5.38 &26.89 &152.98 \\ \hline
30-40\% & 506.16 &51.53 &6.33 &121.56  \\ \hline
40-50\% & 669.40 &125.31 &8.13 &37.70  \\ \hline
50-60\% & 494.09 &140.72 &38.52 &2.75  \\ \hline
 \end{tabular}
\end{table}

Can we understand the centrality dependence of $\eta/s$? Present paradigm is that $\eta/s$ has a minimum, possibly with a cusp, around the critical temperature  $T=T_c$  \cite{Csernai:2006zz}. The centrality dependence of $\eta/s$, as obtained in the present analysis,  is not at variance with the prevailing paradigm. Rather it indicates the increasingly important role of hardronic matter in the development of elliptic flow in peripheral collisions.
As mentioned earlier, viscosity to entropy ratio, as obtained here, is averaged over the space-time. Both the QGP and the hadronic phase contribute to the average. Schematically, one can write,

\begin{equation} \label{eq6}
\frac{\eta}{s} = (1-f_{HAD}) {(\frac{\eta}{s})}^{qgp}(T_{QGP}) + f_{HAD} {(\frac{\eta}{s})}^{had}(T_{HAD})
\end{equation}

\noindent where $f_{HAD}$ is the fraction of the hadronic matter, ${(\frac{\eta}{s})}^{qgp}(T_{QGP})$  is the viscosity of QGP matter at average temperature $T_{QGP}$ and ${(\frac{\eta}{s})}^{had}(T_{HAD})$ is the viscosity
of the hadronic matter at average temperature $T_{HAD}$.
In table.\ref{table3}, we have noted    $f_{HAD}$ at the initial time. In 0-10\% collisions, only $\sim$ 20\% of the matter is in the hadronic phase, the rest ($\sim$80\%) is in the deconfined (or QGP) phase.   Fraction of the hadronic matter increases in peripheral collisions and in 50-60\% centrality Au+Au collisions, at the initial time, $\sim$ 50\% of the total matter is hadronic. Even if elliptic flow   develops early in the evolution,   viscosity of hadronic matter will play important role in the development of the elliptic flow, more so in peripheral collisions. In table.\ref{table3}, we have also noted the spatially averaged initial temperature of the QGP phase  ($T_{QGP}$) and the hadronic phase, ($T_{HAD}$). As expected, $T_{QGP}$ decreases in peripheral collisions,
spatially averaged initial temperature of the hadronic phase, on the other hand is nearly constant, independent of collisions centrality (the
temperature range of the hadronic matter is limited, $150 MeV \leq T_{HAD} \leq 196 MeV$). Since   ${(\eta/s)}^{qgp}$ decreases with decreasing temperature \cite{Csernai:2006zz},  contribution of the QGP phase to the space-time averaged $\eta/s$ will be less in a peripheral collision than in a central collisions.  
Since hadronic fraction increases in peripheral collisions, contribution of the hadronic phase to the space-time averaged $\eta/s$ will increase in peripheral collisions.
Centrality dependence of extracted $\eta/s$ can qualitatively understood as due to increased contribution of hadronic phase and decreased contribution of the QGP phase in peripheral collisions than in a central collisions.

\begin{table}[t] 
\caption{\label{table3} Best fitted $\eta/s$ as a function of collision centrality. Also shown are the fraction of hadronic matter ($f_{HAD}$), spatially averaged temperature of the QGP matter ($\langle T\rangle_{QGP}$), and  the hadronic matter ($\langle T\rangle_{HAD}$) at the initial time $\tau_i$.} 
  \begin{tabular}{|c|c|c|c|c|}\hline
coll. centrality  & $\eta/s$ & $f_{HAD}$  & $\langle T\rangle_{QGP}$ & $\langle T\rangle_{HAD}$    \\
             &     &     & (MeV)     &(MeV)      \\ \hline 
\hline
 0-10\% &$0 \pm 0.03$& 0.20 &298.7 &173.0  \\  \hline
10-20\% &$0.051\pm 0.008$& 0.27 &283.2 &172.8  \\ \hline
20-30\% &$0.087\pm0.004$& 0.31 &267.9 &172.6  \\ \hline
30-40\% &$0.109\pm0.003$& 0.35 &254.7 &172.4   \\ \hline
40-50\% &$0.134\pm 0.004$& 0.41 &239.6 &172.4   \\ \hline
50-60\% &$0.169 \pm 0.005$& 0.49 &222.7 &171.8   \\ \hline
\end{tabular} 
\end{table}

The present estimate $\eta/s$=0-0.17 in 0-60\% centrality range, must be treated with caution. In the present simulations, we have neglected bulk viscosity. Experimental data,   include the effect of bulk viscosity, if there is any.  Neglect of bulk viscosity, will artificially  increase the effect of (shear) viscosity. 
In general, bulk viscosity is an order of magnitude smaller than shear viscosity. But in QGP, it is possible that near the cross-over temperature,
bulk viscosity is large  \cite{Kharzeev:2007wb,Karsch:2007jc}. Effect of bulk viscosity on particle spectra and elliptic flow is studied in \cite{Monnai:2009ad}. It appears that even if small, bulk viscosity can have visible effect on particle spectra and elliptic flow. The present estimate then must be considered as an upper bound on QGP viscosity. 
Also, we have   not considered any systematic error in evaluation of $\eta/s$ due to our uncertain knowledge
about various parameters of the hydrodynamics model.
Systematic error in hydrodynamic evaluation of viscosity could be large. Indeed, in \cite{Chaudhuri:2009uk}, from a simultaneous fit to $\phi$ meson multiplicity, mean $p_T$ and integrated $v_2$, viscosity to entropy ratio was estimated as, $\eta/s=0.07 \pm 0.03 \pm 0.14$, where the 1st error is statistical and the 2nd one is the systematic error. Large ($\sim$ 100\%) systematic error arises due to uncertainty in initial time, initial energy density profile, initial fluid velocity, freeze-out condition, finite accuracy of computer code etc. Even then as noted  in  \cite{Chaudhuri:2009uk}  the source of systematic error is not exhaustive. We expect the systematic error in the present evaluation of viscosity to entropy ratio will be of the same order $\sim$ 100\%.   

\section{Summary and conclusions}

To summarise, we have studied effect of (shear) viscosity  on elliptic flow. To obtain a meaningful comparison between flows in ideal and viscous dynamics, 
the fluid was initialised to reproduce $\phi$ meson multiplicity in 0-5\% Au+Au collisions. The initialisation ensures that  irrespective of fluid viscosity, the final state entropy is the same. Elliptic flow is suppressed in viscous fluid evolution, more viscous is the fluid, more is the suppression. Depending on viscosity ($\eta/s$=0.08-0.16), in central and mid-central ($N_{part} \geq 100$ ) collisions,
integrated $v_2$ is suppressed by 5-15\% only. Suppression of differential $v_2$
is $p_T$ dependent. $v_2(p_T)$ is more suppressed at large $p_T$ than at low $p_T$. Suppression of $v_2(p_T)$ is also centrality dependent, suppression is more in peripheral than in central collisions. Centrality dependence of elliptic flow suppression however reduces at large $p_T$.

We have also compared simulated flow in ideal and viscous dynamics with experiments. PHENIX data \cite{Afanasiev:2009wq} on the differential elliptic flow in the centrality range 0-60\% are best described in
viscous fluid evolution with $\eta/s=0.12\pm 0.005$. However, it was also indicated that the data demand more viscous fluid in more peripheral  collisions.
For example,
in central collisions (0-20\%), PHENIX data are best explained with small viscosity,  $\eta/s \approx$ 0-0.05. In more peripheral collisions, e.g. in 40-60\% collisions, data demand more viscous fluid,  $\eta/s \approx$   0.13-0.17. 
  Centrality dependence of QGP viscosity  can be understood
as due to increased contribution of the hadronic matter in the development of elliptic flow in peripheral collisions.


\begin{thebibliography}{99}
\bibitem{note1} We have checked that with the lattice+HRG EOS, in ideal fluid dynamics, STAR measurements of $\frac{dN^\phi}{dy}$ and $<p_T^\phi>$ in 0-5\% Au+Au collisions are best explained with $T_F$=150 MeV.
\end{thebibliography}
\end{document}